\documentstyle[aps,epsf,prl,preprint]{revtex}
\begin{document}

\title{Convective Fingering of an Autocatalytic Reaction Front}

\author{Michael R. Carey\cite{presentaddress}}
\address{Department of Physics,
University of Toronto, \\60 St. George St., Toronto, Ontario, Canada M5S~1A7}

\author{Stephen W. Morris}
\address{Department of Physics and Erindale College,
University of Toronto, \\60 St. George St., Toronto, Ontario, Canada M5S~1A7}

\author{Paul Kolodner}
\address{AT \& T Bell Laboratories, Murray Hill, New Jersey 07974-0636}

\date{\today}
\maketitle
\begin{abstract}
We report experimental observations of the convection-driven fingering
instability of an iodate-arsenous acid chemical reaction front.  The front
propagated upward in a vertical slab; the thickness of the slab was varied to
control the degree of instability.   We observed the onset and subsequent
nonlinear evolution of the fingers, which were made visible by a {\it p}H
indicator.  We measured the spacing of the fingers during their initial stages
and compared this to the wavelength of the fastest growing linear mode
predicted by the stability analysis of Huang {\it et. al.} [{\it Phys. Rev. E},
{\bf 48}, 4378 (1993), and unpublished].  We find agreement with the thickness
dependence predicted by the theory.
\end{abstract}

Most studies of pattern-forming chemical reactions are carried out in thin
horizontal layers, or in gels, under conditions where the coupling between
chemical concentrations and hydrodynamic effects is suppressed. This coupling
arises because reactions usually modify the density of the solution, giving
rise to buoyancy forces and hydrodynamic flow. This flow then advects the
reacting solution, adding new transport terms to the already complex
reaction-diffusion problem.  In this paper, we study chemical reaction driven
convection in a very simple context, in the autocatalytic reaction of iodate
and arsenous acid in a narrow vertical slot\cite{huang}.  This system may be
regarded as a simplified prototype for the study of hydrodynamic instabilities
in more complex pattern-forming chemical systems, such as the
Belousov-Zhabotinsky reaction\cite{BZ}.

In the absence of hydrodynamic effects, the iodate-arsenous acid
reaction\cite{wellknownRx} produces a single, sharply defined reaction front
which propagates at a speed governed by the diffusion of the iodide
autocatalyst.  When such a front propagates upward in a circular tube, it has
been shown that the solution near the front is unstable to convection when the
tube diameter exceeds a critical value\cite{pojman}.  In this direction of
propagation, the unreacted solution, which is denser, lies above the lighter
reacted solution in a potentially unstable arrangement. For the vertical slot
geometry, which is analytically simpler, there exists a well developed
theory\cite{huang,unboundedtheory,moretheory} of the linear instability of a
flat front. The theory predicts that the initial instability takes the form of
convective rolls which corrugate the front into a series of fingers. In the
limit of sharp fronts, the reaction-diffusion part of the problem can be
reduced to an eikonal relation in which the local front speed depends on the
front curvature\cite{eikonal}.  The relative strength of buoyancy is measured
by a dimensionless parameter $S$, described below, which is proportional to the
cube of the slot thickness. The spacing of the finger pattern results from the
competition between convection, which tends to extend the fingers vertically,
and diffusion, acting via the eikonal relation, which tends to smooth the
front.  More realistic models, which include the full reaction-diffusion
equations needed to describe a finite front thickness, have recently been
proposed\cite{huangprivate}.

A flat front in a laterally extended slot is expected to be unstable to a
periodic chain of convection cells which are analogous to those found in other
simple fluid systems, such as Rayleigh-B\'{e}nard convection\cite{bigreview}.
This instability also bears some analogy to that of a flame front\cite{flames}.
 The objective of the present study is to examine this instability as it grows
out of a flat front, and to compare the spacing of the pattern of fingers to
what is predicted from the linear theory in the thin-front limit.  We also
qualitatively observed the instability after it had become fully developed, in
which the front dynamics is dominated by interactions between fingers. This
highly nonlinear regime is the subject of some recent
simulations\cite{huangprivate}.

\begin{figure}
\epsfxsize =5in
\centerline{\epsffile{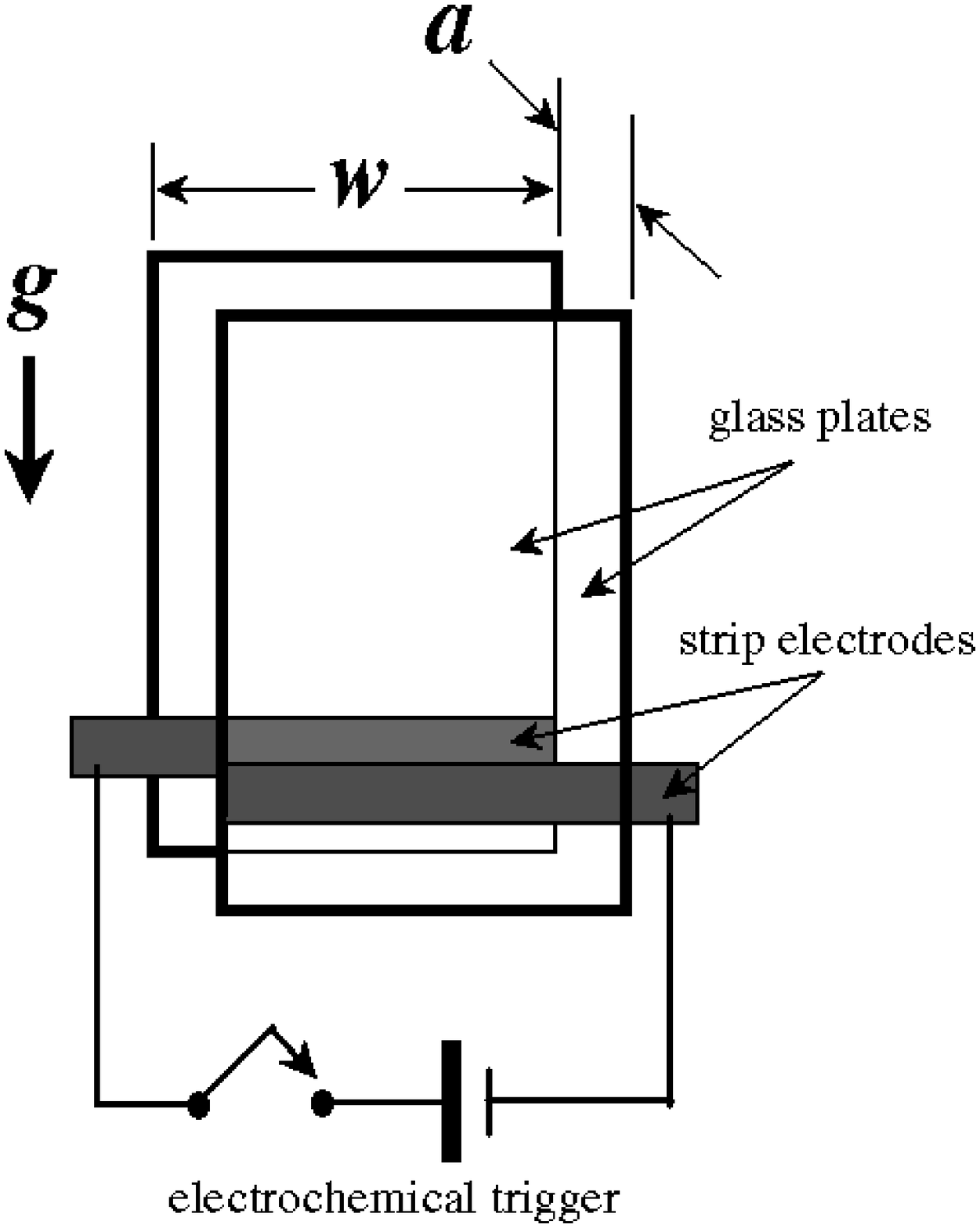}}
\vskip 0.1in
\caption{A schematic of the apparatus. The width $w$ of the cell was $32 mm$.
The thickness $a \sim 1mm$ was varied in the experiment.}
\label{apparatus}
\end{figure}

Fig. \ref{apparatus} shows the experimental arrangement. The slot was formed of
two glass plates held apart by a rubber gasket.  The spacing $a~\sim~1mm$ of
the cell was determined by steel shims outside the gasket, which were not in
contact with the solution.  The width of the slot $w$ was $32 \pm 0.5 mm$.  The
back face of the cell, which was painted white to make the reaction more
visible, was in good thermal contact with a copper block at room temperature,
$23^\circ C$.  The reaction front was initiated by an electrical trigger which
consisted of two thin steel electrodes on opposite sides of the slot which were
in contact with the solution. A DC voltage $\approx 10V$ was applied to the
electrodes for a few seconds to initiate the reaction.  The reaction front,
made visible by an indicator dye, was photographed through the front face of
the cell.  The camera and cell apparatus were mounted on a common stand which
could be oriented with respect to vertical.

The working solution was prepared as follows.  Solid $As_2O_3$ powder was
dissolved in distilled water which had been made strongly basic by the addition
of $KOH$.  When the powder had dissolved, the resulting solution was filtered
and neutralized with concentrated $HCl$.  This was mixed with a solution of
$KIO_3$ and a solution of Congo Red indicator, so that the final concentrations
were $[H_3AsO_3]$~=~0.005M, $[KIO_3]$~=~0.0025M, with 0.0001M of indicator. In
some thinner cells, a higher concentration of indicator was used.  The
unreacted solution was adjusted to have a ${\it p}$H $\approx 6$.  Congo Red
changes from red to blue near ${\it p}$H $\approx 5.2$. With this high initial
${\it p}$H value, the reaction does not start spontaneously, but can be
initiated quickly on an electrode as described above. As the reaction front
passes a point, a sharp transition in color is seen; additional color
gradations in the blue region behind the front served to delineate some flow
structures within the fingers. The front speed in the absence of convection,
measured for downward propagation in narrow capillary tubes, was $c_0 = (3.47
\pm 0.03) \times 10^{-3} mm/s$ for this solution.  This is an order of
magnitude smaller than that observed in previous studies\cite{pojman}, which
used somewhat different concentrations and ${\it p}$H.

We performed runs in cells with various thicknesses $a$ and did not vary the
width $w$, the temperature, or the chemical concentrations. In runs in which
the front was initiated at the top of the slot and propagated downward, the
initial front irregularities grew only slightly during propagation. Fronts
which were initiated at the bottom and propagated upward showed strong
fingering in all but the thinnest cells.  The fingers that formed soon became
nonsinusoidal, taking on a scalloped shape, with broad upper ends and narrow
cusps between fingers on the lower side.  One could clearly see a plume-like
structure within each finger in color gradations of the blue state of the
indicator.  The precise relation of these to the flow is not clear, but they
presumably map varying ${\it p}$H regions which persisted behind the front. The
fingers were reasonably regular initially, but soon after initiation, some
fingers are overtaken and suppressed by their neighbors. There also developed a
tendency for fingers to avoid the lateral edges of the cell. We did not
generally observe tip splitting or mechanisms which nucleated new fingers.  In
the late stages, the front advanced farthest, still carrying interacting
fingers, near the lateral midpoint of the cell.  Eventually, a large central
plume of width $\approx w/2$ developed which was accompanied by a strong
suppression, or even a reversal, of the front advancement near the ends of the
cell.  In the thinnest cells, as discussed below, no fingering was observed in
the early stages, and the only distortion of the front was the development of
this large plume.

\begin{figure}
\epsfxsize =5in
\centerline{\epsffile{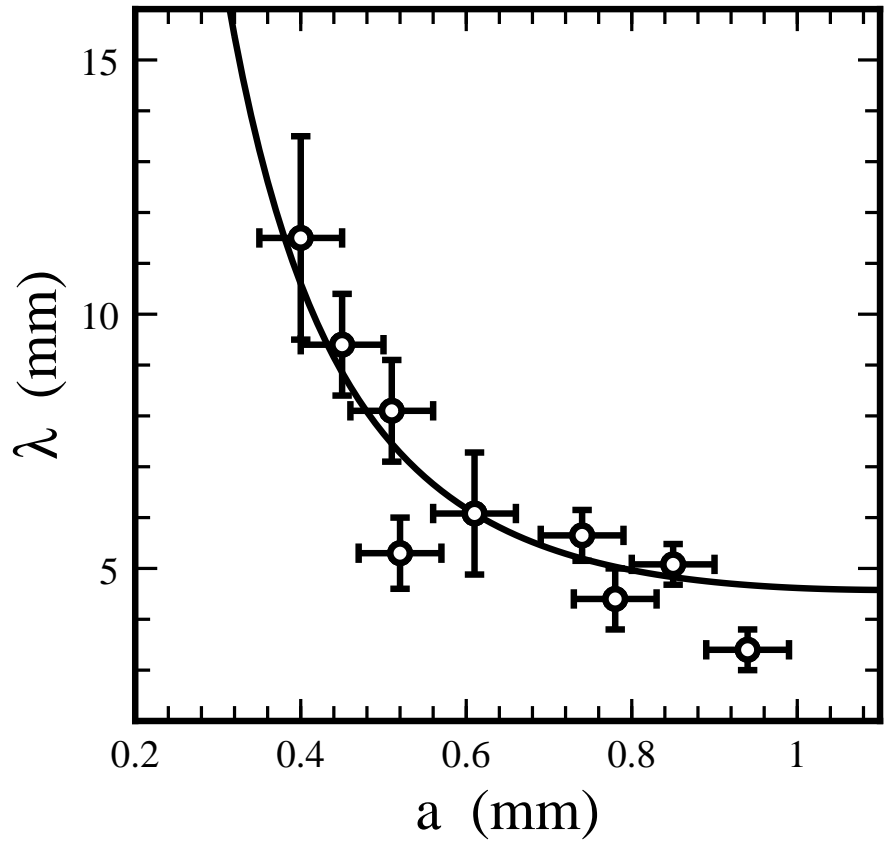}}
\vskip 0.1in
\caption{The pattern wavelength (spacing of the fingers) {\it vs.} the cell
thickness $a$.  The solid line is a fit to the linear theory.}
\label{lambda_vs_a}
\end{figure}

We measured the spacing of the fingers directly from photos, at the earliest
stages of their development for cells of various thickness.  Fig.
\ref{lambda_vs_a} shows a plot of the wavelength $\lambda$ of the fingers,
defined as the mean trough-to-trough spacing, exclusive of end effects, as a
function of the thickness of the cell.  For cells thinner than $0.4$~mm, no
clear fingers were observed before the broad plume reached the top of the cell.
 The main uncertainties in the finger spacing came from the small number of
fingers across the cell (3 - 7), imperfections of the starting conditions,
which tended to be amplified by the instability, and, for thicker cells, the
tendency of the fingers to interact and overtake one another at an early stage.
 However, a clear trend to narrower fingers for thicker cells was evident.  The
solid line in Fig. \ref{lambda_vs_a} is a fit discussed below.

\begin{figure}
\epsfxsize =5in
\centerline{\epsffile{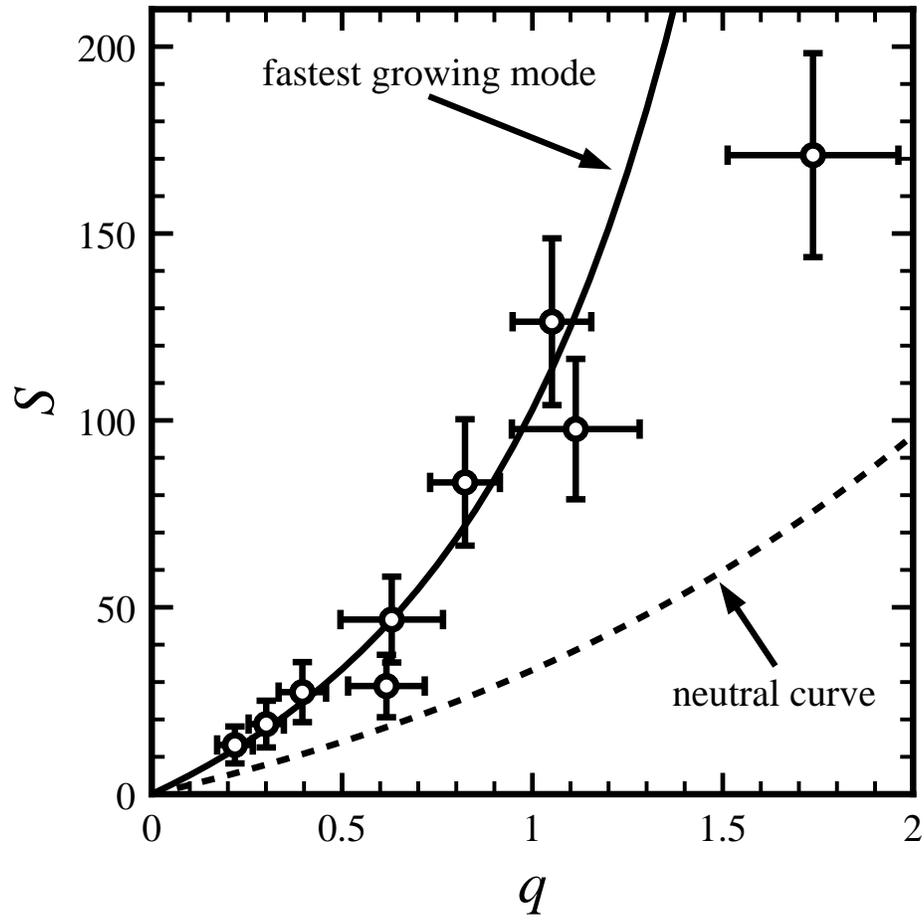}}
\vskip 0.1in
\caption{The dimensionless control parameter $S$ {\it vs.} the wavevector $q$
of the instability, according to the linear theory of Ref.
\protect\cite{huang}.  The data are fit to the fastest growing linear
mode\protect\cite{huangprivate}.}
\label{S_vs_q}
\end{figure}

According to the linear analysis\cite{huang,unboundedtheory,moretheory}, the
dimensionless parameter that controls the instability is given by
\begin{equation}
S = \frac{\delta g a^3 }{\nu D_C},
\label{defineS}
\end{equation}
where $g$ is the acceleration of gravity, $a$ is the cell thickness, $\nu$ is
the kinematic viscosity and $D_C$ the diffusion constant of the autocatalyst.
The density change across the front is parameterized by the dimensionless
density jump $\delta=(\rho_u/\rho_r)-1$, with $\rho_r$ and $\rho_u$ the
densities of the reacted and unreacted solutions. The density change is almost
completely due to concentration change, so that the thermal expansion of the
fluid due to the heat released by the reaction can be
neglected\cite{neglectheat}. For a laterally unbounded slot\cite{huang}, linear
stability analysis predicts that fronts are unstable to convection rolls within
a band of wavenumbers $q$, with $0 \leq q \leq q_c(S)$ for {\it any} $S > 0$.
The critical wavevector $q_c \rightarrow 0$ as $S \rightarrow 0$. The predicted
neutral stability curve $S_c=S(q_c)$, given by Eqn.~40 of Ref.~\cite{huang}, is
shown in Fig. \ref{S_vs_q}.  Physically, this limit means that fronts in wide
slots will be unstable to long wavelength convection, even when the slot is
made very thin ({\it i.e.}, even for small $S$).  For slots of finite width
$w$, one naturally expects a cutoff when the unstable band implies pattern
wavelengths $\lambda \approx w$.  The data shown in Fig. \ref{lambda_vs_a}, for
which $w = 32 \pm 0.5 mm$, are consistent with this expectation; no fingers are
seen in cells with $a < 0.40mm$, a thickness for which only two or three
fingers would fit across the width of the cell.

The theory also predicts the fastest growing linear mode $q_{max}$, which lies
near the midpoint of the unstable band. Fig.\ref{S_vs_q} shows the finger
spacing data on a dimensionless plot of $S$ {\it vs.} $q$, with lengths scaled
by the cell thickness $a$.  We fit the data to the locus of maximum growth
rate\cite{huangprivate} using a one parameter fit of the form $S = k a^3$ {\it
vs.} $q = 2 \pi a/\lambda$, with $k$ adjustable.  The fit minimized the sum of
the absolute deviations, a criterion which is more robust to outliers than the
usual least squares.  We found $k = (2.05 \pm 0.30) \times 10^{11} m^{-3}$. The
fit parameter is simply related to the various parameters in Eqn.
\ref{defineS}. Using the measured value\cite{pojman,unboundedtheory} for $D_C =
2.0 \times 10^{-9} m^2/s$, and taking $\nu$ as the viscosity of water, we find
that the density jump required by Eqn. \ref{defineS}, is $\delta \approx 4
\times 10^{-5}$. This is about 50\% smaller than the
value\cite{pojman,unboundedtheory} previously measured for the isothermal
density change.  The smaller value of this parameter, and our smaller front
speed $c_0$, are probably the result of the different chemical concentrations
and ${\it p}$H of our solution.

The wavelength $\lambda_{max} = 2 \pi a / q_{max}$ corresponding to the fastest
growing linear mode is shown as the solid line in Fig. \ref{lambda_vs_a}.  It
is interesting to note that the linear theory for a laterally unbounded slot
predicts that $\lambda_{max}$ tends to infinity as $a \rightarrow 0$ and to a
constant $\lambda_{\infty} \approx 15(\nu^2 / g \delta)^{1/3}$ as $a
\rightarrow \infty$, and passes through a minimum at intermediate values of
$a$\cite{huangprivate}.  For our solution, this minimum is expected near $a
\approx~1.1~mm$, for which $\lambda_{max} \approx 4.6~mm$, which is
unfortunately just beyond the range of our data.  Observation of this curious
minimum would be a very interesting confirmation of the theory.  In a slot of
finite width, the $\lambda_{max}$ approaches $\lambda_w \sim w$, as $a
\rightarrow 0$, due the constraint that at least one convection roll must fit
across the slot width.  In the late stages of the nonlinear evolution of our
fronts, it is evident that flow eventually develops on the scale of
$\lambda_w$, in the form of a large upflow at the midpoint of the cell, with
downflows near each end.

We have compared our data to the existing theory which is valid in the limit of
zero front thickness\cite{huang}.  In fact, the front has a finite thickness
$d_r \sim D_c/c_0$, which is approximately 0.6~mm for our solution.  Thus $d_r
\sim a$ for most if not all of our data.  While our data do behave as predicted
by the simple theory, we expect that a more accurate model which includes the
effects of finite front thickness will be required for more detailed comparison
to future experiments.

In conclusion, we have observed the fingering instability of a vertically
propagating chemical reaction front in a cell with a slot geometry.  We find
that the spacing of the fingers in the early stages of their development is
consistent with the fastest growing mode according to the linear theory of
convection driven by concentration-induced buoyancy.  We find a long wavelength
cutoff consistent with the effect of the finite lateral extent of the cell.  As
the instability develops, we observed interactions of fingers which tended to
reduce their number, eventually leading to broad flows on the scale of the cell
width.  In future experiments, we plan to extend this study to wider cells,
with better control over the cell geometry, and to cells which are continuously
fed with reactants.  With these arrangements, it should be possible to
precisely study the onset of the instability and the dynamical evolution of the
fingers in the nonlinear regime.

\acknowledgments

We wish to thank Jie Huang and Boyd F. Edwards for communicating the results of
their analysis prior to publication. Some preliminary experiments were
performed by one of us (P.K.) in collaboration with Sibel Bayrakci. M.R.C and
S.W.M. gratefully acknowledge support from the Natural Science and Engineering
Research Council of Canada.

\vfill\eject

\vfill\eject

\vfill\eject

\end{document}